\documentclass[twocolumn,amsmath,amssymb,showpacs,aps,prb,superscriptaddress]{revtex4-1}

\usepackage[colorlinks=true,citecolor=blue,linkcolor=blue]{hyperref}
\usepackage{mathrsfs}
\usepackage{mathtools}
\usepackage{xfrac}
\usepackage{graphicx}
\usepackage{dcolumn}
\usepackage{bm}
\usepackage{color}
\usepackage{tabularx}
\usepackage{amsmath}

\newcommand{\sign}{\text{sign}}

\newcommand{\bn}{{\bf n}}

\newcommand{\bk}{{\bf k}}
\newcommand{\bq}{{\bf q}}

\newcommand{\bd}{{\bf d}}

\newcommand{\bQ}{{\bf Q}}

\newcommand{\br}{{\bf r}}

\newcommand{\ve}{{\varepsilon}}

\newcommand{\Tr}{\text{Tr}}

\newcommand{\be}{\begin{equation}}
\newcommand{\ee}{\end{equation}}
\newcommand{\beq}{\begin{eqnarray}}
\newcommand{\eeq}{\end{eqnarray}}
\newcommand{\bea}{\begin{align}}
\newcommand{\eea}{\end{align}}
\newcommand{\beqq}{\begin{eqnarray*}}
\newcommand{\eeqq}{\end{eqnarray*}}

\newcommand{\up}{\uparrow}
\newcommand{\down}{\downarrow}

\begin{document}

\begin{titlepage}

\title{Nematic superconductivity stabilized by density wave fluctuations: \\
Possible application to twisted bilayer graphene}

\author{Vladyslav Kozii}
\affiliation{Department of Physics, Massachusetts Institute of Technology,
Cambridge, MA 02139, USA}

\author{Hiroki Isobe}
\affiliation{Department of Physics, Massachusetts Institute of Technology,
Cambridge, MA 02139, USA}

\author{J\"orn W. F. Venderbos}
\affiliation{Department of Physics and Astronomy, University of Pennsylvania, Philadelphia, Pennsylvania 19104, USA}
\affiliation{Department of Chemistry, University of Pennsylvania, Philadelphia, Pennsylvania 19104, USA}

\author{Liang Fu}
\affiliation{Department of Physics, Massachusetts Institute of Technology,
Cambridge, MA 02139, USA}

\begin{abstract}
Nematic superconductors possess unconventional superconducting order parameters that spontaneously break rotational symmetry of the underlying crystal. In this work we propose a mechanism for nematic superconductivity stabilized by strong density wave fluctuations in two dimensions. While the weak-coupling theory finds the fully gapped chiral state to be energetically stable, we show that strong density wave fluctuations result in an additional contribution to the free energy of a superconductor with multicomponent order parameters, which generally favors nematic superconductivity. Our theory sheds light on the recent observation of rotational symmetry breaking in the superconducting state of twisted bilayer graphene~\cite{Cao12018, Pablounpublished}.
\end{abstract}

\pacs{74.20.Rp, 74.20.Mn, 74.45.+c}

\maketitle

\draft

\vspace{2mm}

\end{titlepage}

\section{Introduction}
Unconventional superconductors can have multicomponent superconducting order parameters which transform in a multidimensional representation of the crystal symmetry group. In such cases, additional symmetries besides the $U(1)$ gauge symmetry --- such as time-reversal or rotation symmetry --- are broken in the superconducting state. Superconductors which break rotation symmetry can be called nematic superconductors (NSC), in analogy with rotational symmetry breaking in liquid crystals, whereas superconductors which break time-reversal symmetry are known as chiral superconductors. Nematic and chiral superconducting order parameters that belong to the same multiplet are degenerate at the superconducting transition temperature, while this degeneracy is lifted at lower temperature.

Recently, NSC have attracted a lot of attention following the discovery of rotation symmetry breaking in superconducting states of  doped topological insulators $\text{Bi}_2 \text{Se}_3$ 
in  Knight shift~\cite{G.Q.Zheng2016}, upper critical field~\cite{Maeno2017,Willaetal2018,deVisser2016,Nikitin2016,Lortz2017}, specific heat~\cite{Maeno2017,Willaetal2018}, magnetic torque~\cite{LuLi2017} and STM~\cite{TaoYanLiu2018} measurements. Importantly, no signatures of rotational symmetry breaking were found in the normal state, indicating that nematicity is a property of the superconducting state itself.
The observed features are consistent with a nematic superconductor with a two-component odd-parity SC order parameter ~\cite{Fu2015, FuBerg}.

The studies of NSC have mainly focused on strongly spin-orbit-coupled 3D materials and have considered odd-parity pairings~\cite{Venderbos2016,Martin2017,Zhang2015,deVisser2018,Snezhko2016,Welp2017,Smylie2017,VenderbosHc22016,Babaev2017,Martin2017vortex,Schmalian2018,Yao2018,Sauls2018}.
As shown by weak-coupling approach~\cite{Venderbos2016,Martin2017}, in the presence of strong spin-orbit coupling and odd-parity pairing, the nematic superconducting state can be energetically more favorable than the chiral one due to the difference in their gap structures.
In contrast, for two-dimensional (2D) systems without spin-orbit coupling, NSC is not expected from the gap structure: In 2D, the chiral $p+ip$ or $d+id$ SC states generally have a full superconducting gap on the Fermi surface, whereas the nematic $p_x$ ($p_y$) or $d_{x^2-y^2}$ ($d_{xy}$) states have point nodes. As a result, the chiral state has a lower energy compared to the nematic state within a weak-coupling treatment~\cite{Annica2007,ChengPRB2010,Chirolli,Nandkishoreetal.,NandkishorePRB2014}.

In this work, we propose a mechanism for  $p$- or $d$-wave nematic superconductivity in 2D systems with hexagonal symmetry  $D_6$. We focus on the vicinity of the superconducting transition temperature, which allows us to treat the problem within the Ginzburg-Landau (GL) theory. By going beyond the weak-coupling approach which only takes into account the energy of Bogoliubov quasiparticles, we show that sufficiently strong fluctuations of a density wave order  stabilize  the NSC. The energy of such fluctuations is affected by  the presence of a pairing potential and can thus  distinguish between different SC states. Usually, the corresponding contribution is small compared to the weak-coupling term; however, it becomes more significant as the strength of fluctuations grows. This effect is known as a feedback mechanism and was originally proposed by Anderson and Brinkman to explain the stability of nodal $A$-phase in superfluid He-3 due to strong ferromagnetic fluctuations \cite{andersonbrinkman,brinkmansereneanderson,leggett}.

We show that strong density wave fluctuations generically favor nematic superconductivity in 2D. Our results are largely independent of microscopic details of such fluctuations. We find that the nematic $d$-wave state is  stabilized more significantly by charge density wave (CDW) fluctuations, while the feedback contribution from spin density wave (SDW) is partially suppressed by the destructive interference in a coherence factor. Interestingly, the conclusion is dual in a case of two-component spin-triplet superconductivity, i.e., nematic $p$-wave SC is more stabilized by SDW.  We emphasize that our analysis is valid not far from the superconducting transition point, where GL theory is applicable.

Our work is motivated by recent experiments on twisted bilayer graphene (TBG), which observed superconductivity and strongly correlated insulating state at 'magic' angle $\theta \approx 1.1^{\circ}$ ~\cite{Cao12018,Cao22018,Dean2018}. The mechanism for superconductivity in TBG and the pairing symmetry are subject to intense theoretical study~\cite{IsobeYuanFu,Laksonoa2018,Balents12018,YouVishwanath2018,ZhuXiangZhang2018,Volovik2018,Senthil12018,Phillips2018,Phillips22018,Baskaran2018,Kivelson2018,Trivedi2018,Yang2018,Walet2018,Kaxiras2018,Stauber2018,Sherkunov2018,Bernevig22018,Lin2018,Fernandes2018,
Wu2018,Martin2018,LinNandkishore2018,KurokiPRB2018,Han2018,TangYangZhangWang2018,ChoiChoi2018}. In particular, Ref.~\onlinecite{IsobeYuanFu} showed that due to the Fermi surface nesting and the proximity to Van Hove singularity, unconventional superconductivity and density wave emerge as the two leading instabilities driven by Coulomb interaction. The most divergent superconducting instabilities are found in the two-component  $p$-wave and $d$-wave superconducting channels, which are nearly degenerate when the intervalley exchange interaction is small. Related results on superconductivity from density wave/antiferromagnetic fluctuations in graphene superlattices appear in Refs.~\onlinecite{Balents12018,YouVishwanath2018,ZhuXiangZhang2018} (which propose chiral $p+ip$ and $d+id$ pairings).
We thus expect that our strong-coupling theory of nematic superconductivity from density wave fluctuations is  directly applicable to TBG. The strength of density wave fluctuations, which are important for our theory, may be enhanced by, e.g., changing electron density or tuning the twist angle and the interlayer spacing. We predict that one can observe the transition between chiral and nematic $p$/$d$-wave superconducting states upon varying these parameters.

Our result sheds light on the  most recent measurements of in-plane upper critical magnetic field in TBG, which shows a pronounced two-fold anisotropy revealing the breaking of rotational symmetry~\cite{Pablounpublished}. This experimental finding suggests the possibility of nematic $p$-wave or $d$-wave superconductivity, instead of chiral $p+ip$ or $d+id$ states which are isotropic. We emphasize that in our study nematicity is an intrinsic property of the anisotropic superconducting state, and we assume there is no primary electronic order that breaks rotation symmetry in the normal state. This should be contrasted with the scenario considered in Ref.~\onlinecite{Kivelson2018}, where the anisotropy of the superconducting state originates from nematic orbital order that onsets at high temperature in the normal metal. Furthermore, the nematic superconducting state studied in Ref.~\onlinecite{Kivelson2018} is fully gapped and breaks time-reversal state, while the $p$- or $d$-wave nematic superconductor studied in our work is time-reversal-symmetric and has point nodes in the gap structure. This can be directly probed in future experiments, including thermal transport and tunneling spectroscopy.


\begin{figure}
\centerline{\includegraphics[width=.49\textwidth]{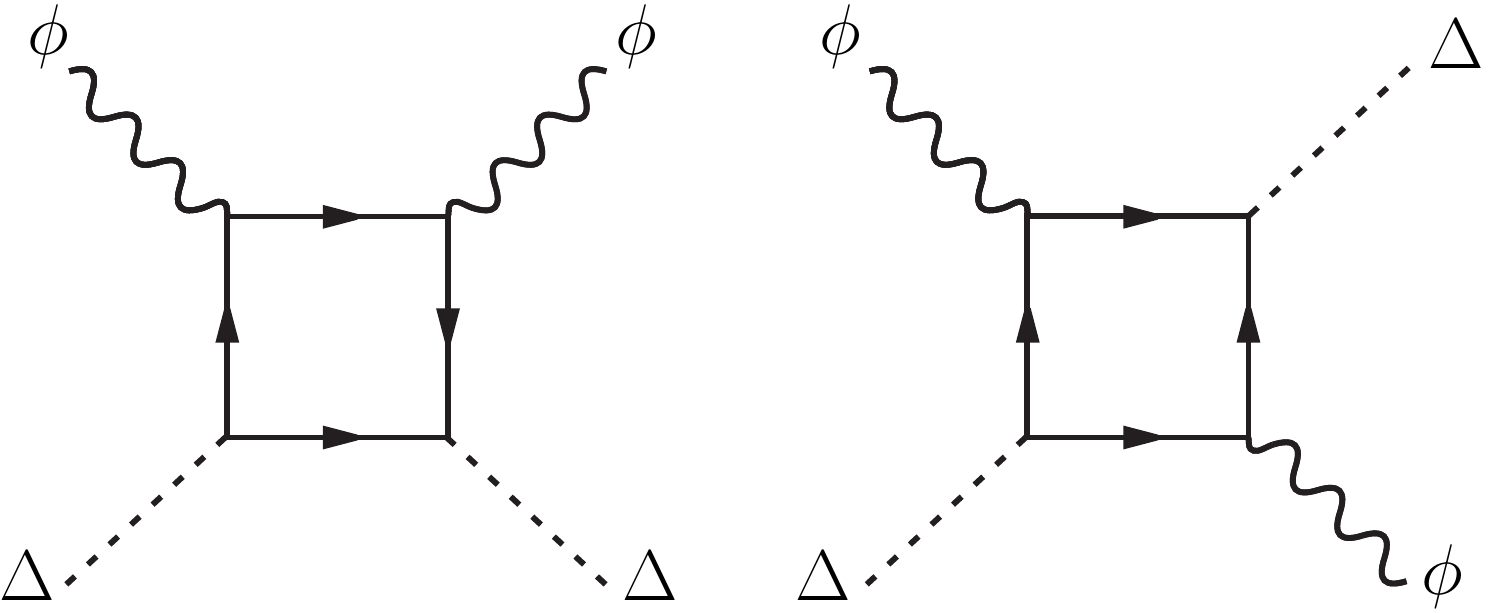}}
\caption{Two lowest-order diagrams describing coupling between CDW fluctuations $\phi_i$ and SC order parameter $\Delta$, see Eq.~(\ref{Eq:deltaFphiDelta}).  These diagrams are sufficient provided the system is close to the superconducting transition ($\Delta$ is small) and the CDW fluctuations are massive.}
\label{Fig:diagrams}
\end{figure}

\section{CDW modes coupled to SC}
To understand the essential physics of strong density wave fluctuations coupled to SC, we begin by considering a phenomenological GL theory. In GL theory the free energy $F_{\text{SC}}$ of the superconductor is expanded up to fourth order in the two-component (spin-singlet) $d$-wave order parameter  $\hat \Delta = \Delta \cdot (d_1, d_2)$, i.e., $F_{\text{SC}}=F^{(2)}_{\text{SC}} +F^{(4)}_{\text{SC}} $, with $F^{(4)}_{\text{SC}}$  for the hexagonal systems given by

\be
F^{(4)}_{\text{SC}} = \alpha_1 \Delta^4(|d_1|^2 + |d_2|^2)^2 + \alpha_2 \Delta^4 |d_1^2 + d_2^2|^2, \label{Eq:FSC}
\ee
where $\alpha_{1,2}$ are the GL expansion coefficients. The sign of $\alpha_2$ determines the SC state below $T_c$:
If $\alpha_2 > 0$, chiral superconducting state has lower energy, $(d_1, d_2) \sim (1, \pm i).$ This state breaks time-reversal symmetry, since $d_i \to d_i^*$ under time reversal, and is characterized by full pairing gap on the entire Fermi surface. In contrast, if $\alpha_2 < 0$, the order parameter is real and given by  $(d_1, d_2) \sim (\cos \theta, \sin \theta).$ This state defines a nematic superconductor, owing its name to the nonzero subsidiary nematic order
\be
(N_1, N_2) = (|d_1|^2 - |d_2|^2, d_1^*d_2 + d_1 d_2^*),  \label{Eq:nematic-SC}
\ee
which transforms as a nematic director; this state has nodes in the excitation spectrum. As shown below, calculating $\alpha_{1,2}$ within weak-coupling gives $\alpha_2 > 0$, selecting the chiral state.

Next, we introduce the coupling to density wave fluctuations. For the sake of definiteness, we consider CDW fluctuations, but note that the argument is similar for SDW fluctuations. In hexagonal systems CDW order is described by a three-component complex order parameter $\boldsymbol{\phi} = (\phi_1,\phi_2,\phi_3)$, where the fields $\phi_i$ correspond to CDW modes at ordering wave vectors $\bQ_i$. These three wave vectors are related by sixfold rotation.  To the lowest order, the coupling of the SC order parameter $\bd$ to the CDW modes $\boldsymbol{\phi}$, shown diagrammatically in Fig.~\ref{Fig:diagrams}, can be expressed as

\be
F_{\phi-\Delta} = \beta_1 |\boldsymbol{\phi}|^2 |\bd|^2 + \beta_2(P_1 N_1 + P_2 N_2). \label{Eq:deltaFphiDelta}
\ee
Here $(P_1, P_2) = (2 |\phi_1|^2 - |\phi_2|^2 - |\phi_3|^2, \sqrt{3}(|\phi_2|^2 - |\phi_3|^2))$ is a subsidiary nematic order parameter quadratic in the fields $\phi_i$ and describes anisotropic CDW fluctuations which transform as partners under rotations, and $\bd = (d_1, d_2)$. Since it has the same symmetry as $(N_1, N_2)$ in \eqref{Eq:nematic-SC} it couples linearly.


We further assume that the fields $\phi_i$ are massive and can be described by a Gaussian contribution $F_\phi$, the precise form of which is immaterial for present purpose. The free energy of the superconductor coupled to the CDW fluctuations can thus be expressed as $F= F_\Delta + F_\phi+F_{\phi-\Delta}$. Since the fields $\phi_i$ are massive they can be integrated out, which leads to an effective free energy for the superconductor given by
$ F_{\text{SC}}=F_\Delta  + \delta F_\Delta $; at fourth order, the correction $\delta F^{(4)}_\Delta $ is given by
\be
\delta F^{(4)}_\Delta  \sim -\beta_1^2 |\bd|^4 - 2\beta_2^2(N_1^2 + N_2^2).\label{Eq:FphiDeltaaveraged}
\ee
Using the identity $N_1^2 + N_2^2 = |d_1^2 + d_2^2|^2$, we observe that \eqref{Eq:FphiDeltaaveraged} implies a lowering of the energy of the nematic superconducting state relative to the chiral state. This effect is enhanced as the fluctuations become stronger, thus exceeding the weak-coupling or any other fourth-order contribution and eventually leading to NSC. Remarkably, the argument leading to Eq. \eqref{Eq:FphiDeltaaveraged} is general and does not rely on the nature of the fluctuating field.
In particular, as mentioned, it also applies to SDW fluctuations, which can be described by a vectorial order parameter $\vec{ \boldsymbol{\phi}} = (\vec \phi_1,\vec \phi_2,\vec \phi_3)$.
Finally, a similar argument was applied to demonstrate the existence of $s+d$-wave SC state in the presence of nematic fluctuations in systems with tetragonal symmetry~\cite{FernandesMillis}.

\begin{figure}
\centerline{\includegraphics[width=1.\columnwidth]{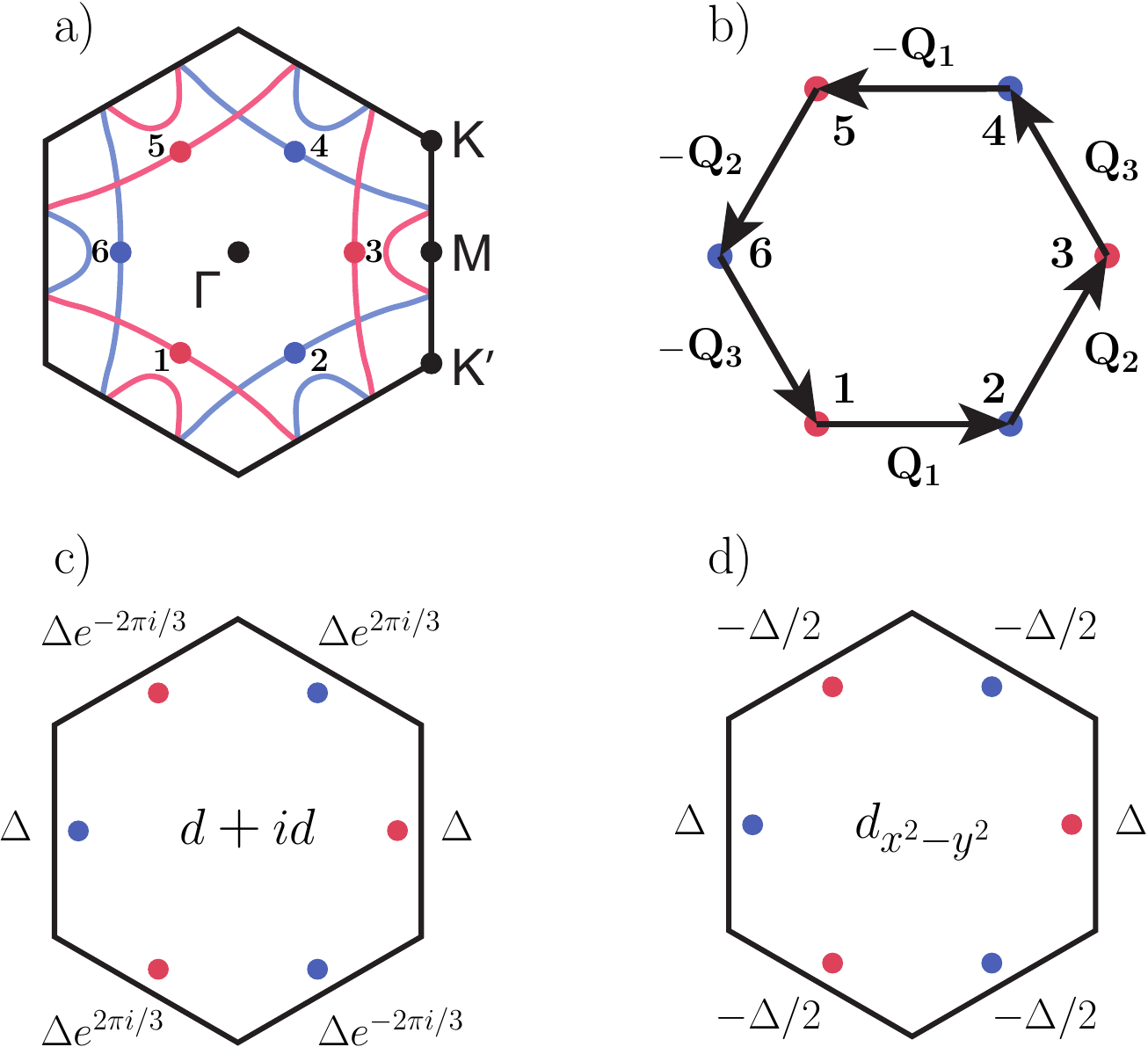}}
\caption{(a) Fermi surface of twisted bilayer graphene slightly away from Van Hove singularity. Blue and red parts originate from different valleys. (b) Hot spots are  connected by six inequivalent CDW/SDW wavevectors $\pm \bQ_i$, related by six-fold rotations. For concreteness, we assume that $\pm \bQ_i$ connect adjacent hot spots.  (c) Time-reversal-breaking chiral superconductivity is realized if $\alpha_2 > 0$, see Eq.~(\ref{Eq:FSC}), while (d) the rotational symmetry-breaking nematic state has lower energy provided $\alpha_2 < 0$. }
\label{Fig:hotspots}
\end{figure}

\section{General model for CDW fluctuations and SC}
While the above argument is physically compelling and correctly captures the  physical mechanism of fluctuation-induced NSC, it is based on a simplified approach which neglects the contribution of modes with nonzero momentum or frequency. To develop a theory of NSC which takes this into account, we now consider a more general model for a two-component $d$-wave superconductor in the presence of CDW fluctuations. The Hamiltonian of such a system is given by $H = H_\psi+ H_\phi + H_{\psi-\Delta}  + H_{\psi -\phi}$, where

\be
H_\psi = \sum_{\bk \alpha} \xi_{\bk} \psi^\dagger_{\bk \alpha}  \psi_{\bk \alpha}, \quad H_\phi = \frac12\sum_{\bq} \tilde V_0^{-1}(\bq) \phi_\bq \phi_{-\bq}   \label{Eq:H-A}
\ee
describe the normal state electronic excitations $ \psi_{\bk \alpha}$ with dispersion $\xi_{\bk}$ and spin $ \alpha=\up,\down$, and the (bosonic) CDW fluctuations $\phi_\bq$ governed by the bare propagator $\tilde V_0(\bq)$, respectively. The propagator $\tilde V_0(\bq)$ is peaked at the six symmetry-related CDW ordering vectors $\pm \bQ_{i=1,2,3}$. The coupling of the fermions to the superconducting pair potential and the CDW fluctuations is given by

\begin{align}
&H_{\psi-\Delta} = \frac12\sum_{\bk}\left(\psi^\dagger_{\bk \uparrow} \psi^\dagger_{-\bk \downarrow} - \psi^\dagger_{\bk \downarrow} \psi^\dagger_{-\bk \uparrow} \right) \Delta_{\bk} + \text{H.c.}, \nonumber \\
&H_{\psi-\phi} = \lambda  \sum_{\bk, \bq} \psi_{\bk + \bq \alpha}^\dagger \psi_{ \bk \alpha}  \phi_\bq, \label{Eq:H-B}
\end{align}
respectively.  On the Fermi surface, the pairing potential of the two-component $d$-wave SC is given by $\Delta_{\bk} = \Delta[2 d_1 \hat k_x \hat k_y + d_2 (\hat k_x^2 - \hat k_y^2)]$, where, again, $\Delta$ is the overall pairing strength and $\bd = (d_1,d_2)$, which satisfies $|\bd|^2=1$,  captures the structure of the two-component order parameter. This form of the pairing potential corresponds to the $E_2$  representation of the point group $D_6$.

The Hamiltonian $H$ of Eqs.~\eqref{Eq:H-A} and \eqref{Eq:H-B} defines a general model for a $d$-wave superconductor coupled to CDW fluctuations. To demonstrate how the (gapped) fluctuations can induce NSC via the so-called feedback mechanism, we proceed in two main steps: First, we integrate out the fermions and then the fluctuation fields $\phi_\bq$. In this way, we obtain an effective free energy functional for the superconducting order parameter which includes the effect of CDW fluctuations and renormalizes the weak-coupling result. The latter is directly obtained from $H$ by neglecting the effect of CDW fluctuations altogether. More precisely, within weak-coupling the BCS free energy of the superconductor is given by $F_{\Delta} = -T \ln \left[\Tr \exp(-H_0/T) \right]$, where $H_0 = H_\psi + H_{\psi-\Delta}$. After straightforward evaluation, we find $F_{\Delta}$ in hexagonal systems up to fourth order as (see Appendix~\ref{SMSec:CDW} for details)

\be
F_{\Delta} = r \Delta^2 |\bd|^2  + K_0 \Delta^4 \left(2 |\bd|^4 + |\bd^2|^2\right),  \label{Eq:F0}
\ee
where $ r\sim (T-T_c)$, $K_0 = (T/16) \sum_{\omega_n, \bk} ( \omega_n^2 + \xi_{\bk}^2 )^{-2}$, and $\omega_n = \pi T (2 n+1)$ are fermionic Matsubara frequencies. Since $K_0 > 0$, we find that within weak-coupling the chiral state indeed has lower energy below $T_c$, in agreement with Ref.~\onlinecite{Nandkishoreetal.}. This applies very generally to systems with hexagonal symmetry and does not depend on microscopic details of Fermi surface.

\section{Fluctuation-induced NSC}
We now proceed to calculate the correction to the weak-coupling free energy originating from the CDW fluctuations. As an intermediate step, we consider the normal state electronic structure described by $\xi_\bk$ of \eqref{Eq:H-A} in more detail. At low energies, the most important electronic excitations are located in the vicinity of those points on the Fermi surface which are connected by the CDW ordering vectors, the so-called hot spots. The hexagonal symmetry dictates that there are six such hot spots. Following the results of Ref.~\onlinecite{IsobeYuanFu}, we focus on CDWs with wavevectors that connect all adjacent hot spots, as shown in Fig.~\ref{Fig:hotspots}(b), though this assumption is not important for our analysis. This hot spot model, which introduces six flavors of low-energy fermions $\psi_{i \bk \alpha}$ ($i=1,\ldots,6$) with corresponding $\xi_{i \bk}$, also establishes a natural connection with TBG~\cite{IsobeYuanFu,Laksonoa2018}. Note that within the hot spot model all momenta are measured with respect to the hot spots.

To calculate the feedback correction, we integrate out the CDW fluctuations, which we assume to be massive, and determine their contribution to the free energy. Importantly, this contribution depends on the SC order parameter and can be expanded in $\Delta$ to obtain renormalized GL coefficients. Since the full calculation is tedious but straightforward, we discuss only  the final result and present the details in Appendix~\ref{SMSec:CDW}. We find the correction to the free energy due to CDW fluctuations as

\be
\delta F_{\Delta} = - T\Tr \big( \hat V \delta \hat\chi  \big) - \frac{T}2\Tr \big( \hat V \delta \hat\chi \big)^2+\ldots, \label{Eq:Fphi}
\ee
where $\hat V$ is an effective propagator of the CDW fluctuations, and $\delta \hat \chi$ is the correction to the CDW susceptibility due to the coupling to $\Delta$. Note that in \eqref{Eq:Fphi} $\Tr$ implies summation over frequencies, momenta, as well as patches.  Near $T_c$, $\delta \hat \chi$ can be expanded in powers of $\Delta$, with the lowest-order term proportional to $\Delta^2$. This term is diagrammatically shown in Fig.~\ref{Fig:diagrams} and in the limit of zero frequency and momentum has exactly the form of Eq.~\eqref{Eq:deltaFphiDelta}. As a result, the term in Eq.~\eqref{Eq:Fphi} proportional to $\Tr ( \hat V \delta \hat\chi )^2$ gives rise to a quartic correction $\delta F_{\Delta}^{(4)}$ to the free energy of the superconductor and shifts the energetic balance towards the nematic state,  in agreement with Eq.~\eqref{Eq:FphiDeltaaveraged}. As we discuss in Appendix~\ref{SMSec:CDW}, the effect of the term proportional to $\Tr ( \hat V \delta \hat\chi )$ is less important and can be neglected as the strength of fluctuations increases.

Remarkably, we find that $\delta F_{\Delta}$ always favors nematic SC, irrespective of the precise form of $\hat V(\Omega, \bq)$ or $\xi_\bk$, provided CDW fluctuations are sufficiently strong. This result solely relies on the form of $\delta \hat \chi$, and is a direct generalization of Eq.~(\ref{Eq:FphiDeltaaveraged}) for the case when CDW modes with nonzero $\Omega$ and $q$ are taken into account. Specifically,  for the model described by Eqs.~\eqref{Eq:H-A} and \eqref{Eq:H-B}, the leading contribution to the fourth-order
free energy feedback correction $\delta F_{\Delta}^{(4)}$ reads as
\begin{multline}
\delta F_{\Delta}^{(4)} = -\frac{3 T^3 (\lambda  \Delta)^4}{2} \left[ (Y_1 + 8 Y_2 + 8 Y_3 + 8 Y_4)|\bd|^4 \right. \\ \left. + 2(Y_1 + 2 Y_2 + 2 Y_3 -  Y_4) |\bd^2|^2  \right], \label{Eq:Fphigeneral}
\end{multline}
with
\begin{align}
&Y_1 \equiv \sum_{\bq, \Omega_m} V^2(\bq) K_2^2(\bq), \, Y_3 \equiv \sum_{\bq, \Omega_m} V^2(\bq) K_1(\bq) K_2(\bq), \nonumber \\ &Y_2 \equiv \sum_{\bq, \Omega_m} V^2(\bq) K_1^2(\bq) , \, Y_4 \equiv \sum_{\bq, \Omega_m} V^2(\bq) K_1(\bq) K_1(-M \bq), \label{Eq:Y1-Y4}
\end{align}
and the functions $K_1$ and $K_2$ are defined as
\begin{align}
&K_1(\bq,\Omega_m) \equiv \sum_{\bk, \omega_n} \frac{\omega_n (\omega_n + \Omega_m) - \xi_{1\bk} \xi_{2 \bk - \bq}  }{ \left[\omega_n^2 + \xi_{1\bk}^2 \right]^2\left[ (\omega_n+ \Omega_m)^2 + \xi_{2 \bk - \bq}^2 \right]}, \nonumber \\
&K_2(\bq, \Omega_m) \equiv \sum_{\bk, \omega_n} \frac{1}{ \left[ \omega_n^2 + \xi_{1\bk}^2 \right]\left[ (\omega_n+ \Omega_m)^2 + \xi_{2 \bk - \bq}^2 \right]}. \label{Eq:K1K2}
\end{align}
Here $\xi_{1,2 \bk}$ are the dispersions of the hot spot fermions, see Fig.~\ref{Fig:hotspots}, which are related by the mirror symmetry $M$ as $\xi_{2 \bk} = \xi_{1 M \bk}$. $\omega_n = \pi T (2n + 1)$ and $\Omega_m = 2\pi T m$ are fermionic and bosonic Matsubara frequencies, respectively, and we have suppressed $\Omega_m$ in \eqref{Eq:Y1-Y4} for brevity.
Since $V(\Omega,\bq)$ in \eqref{Eq:Y1-Y4} is the CDW propagator within the hot spot model it is peaked at $\bq =0$; $V(\Omega,\bq)$ is related to the (effective) propagator of the full model  $\tilde V$ as $V(\Omega, \bq) \equiv \tilde V(\Omega, {\bf Q}_1 + \bq)$, where ${\bf Q}_1$ connects hot spots 1 and 2, see Fig.~\ref{Fig:hotspots}.


Our claim that the feedback effect of CDW fluctuations always favors NSC now follows from Eq.~\eqref{Eq:Fphigeneral}: It is easily demonstrated that the coefficient of $|d_1^2 + d_2^2|^2$ is always positive, i.e., $Y_1 + 2 Y_2 + 2Y_3 -  Y_4 > 0$, thus proving our statement. Importantly, Eq.~\eqref{Eq:Fphigeneral} does not require any assumptions about the explicit form of $V(\Omega, \bq)$ or $\xi_{i\bk}$, and only relies on the (rotation and mirror) symmetries relating the hot spots. Furthermore,
when $V(\Omega, q)$ is strongly peaked at $\Omega = q = 0$, we exactly recover Eqs.~\eqref{Eq:deltaFphiDelta}-\eqref{Eq:FphiDeltaaveraged} with
\begin{align}
&\beta_1 = T (\lambda  T \Delta)^2 \left[ 4 K_1(0,0) + K_2(0,0)  \right], \nonumber \\ &\beta_2 = T (\lambda T \Delta)^2 \left[ K_1(0,0) + K_2(0,0)  \right]. \label{Eq:beta12}
\end{align}
The overall prefactor in Eq.~\eqref{Eq:FphiDeltaaveraged} is proportional to $\sum_{\bq} V^2(0,q),$ implying that the feedback contribution becomes more significant as the strength of CDW fluctuations grows. As such, Eq.~\eqref{Eq:Fphigeneral} in combination with \eqref{Eq:FphiDeltaaveraged} represents the central result of this paper.

Finally, we consider a specific model for twisted bilayer graphene to exemplify our results. We assume that the effective normal-state CDW propagator is given by  $V(q) = \chi_0/c^2(q_0^2 + q^2)$.  We use the Fermi surface reproduced from the band structure calculation for TBG with filling factor close to the filling of two electrons/holes per supercell~\cite{KimNanoLett}, see Fig.~\ref{Fig:hotspots}(a). The dispersion near the hot spots can be  approximated as $\xi_{i\bk} = v (\bk \cdot {\bf n}_i),$ where ${\bf n}_i$ is the unit vector in the direction $\Gamma \text{M}_i$. The feedback correction to free energy then equals
\be
\delta F_{\Delta}^{(4)} = - \frac{13.46}{ (2\pi)^4}\left(\frac{\chi_0 T \lambda^2}{ q_0 v^2 c^2} \right)^2 \frac{\Delta^4}{T^3} \left( 1.16 |\bd|^4 + |\bd^2|^2  \right). \label{Eq:answer2}
\ee

As the strength of the fluctuations increases, i.e., as $q_0$ becomes smaller, this correction becomes more significant, eventually exceeding the weak-coupling contribution and leading to NSC. We emphasize that while the numerical prefactors in \eqref{Eq:answer2} depend on the particular model chosen, the results given by Eqs.~\eqref{Eq:Fphigeneral}--\eqref{Eq:K1K2} were derived without specifying the normal-state dispersion $\xi_{i\bk}$ and normal-state CDW propagator $V(\Omega, \bq)$, and thus apply very generally.
In particular, it can be used in the case when the Fermi energy is close to Van Hove singularity, and the hot spot dispersion (in proper coordinates) is given by $\xi_{\bk} = A k_x^2 - B k_y^2$ with some constants $A, B >0$. In the latter case, we reach the same qualitative conclusion that sufficiently strong density wave fluctuations stabilize NSC (see Appendix~\ref{SMSec:specificmode2} for details).

The analysis for the case of strong SDW fluctuations in a $d$-wave superconductor is similar. The important difference, however, is the relative minus sign between two diagrams in Fig.~\ref{Fig:diagrams}. This leads to the destructive interference for the coherence factor in the expression for $\delta \hat \chi$~\cite{Tinkham}. It is straightforward to show that all results for  SDW fluctuations, apart from a possible overall numerical prefactor, can be obtained from the CDW case simply by changing $K_2(\bq, \Omega) \to -K_2(\bq, \Omega)$, which leads to the partial (but not complete) suppression of the feedback contribution to free energy (see Appendix~\ref{SMSec:SDW}).

\section{Conclusion}
In conclusion, by going beyond weak-coupling and including fluctuations of a density wave order we have presented a general mechanism for nematic multicomponent superconductivity in 2D. The theory we develop can be directly applied to twisted bilayer graphene, where density wave fluctuations are strong due to Fermi surface nesting and intertwined with $d$/$p$-wave superconductivity \cite{IsobeYuanFu}. Together with the recent observation of the upper critical magnetic field anisotropy in twisted bilayer graphene~\cite{Pablounpublished}, this serves as a main experimental motivation for our work.

As mentioned in the introduction, in our theory the onset of nematicity is tied to pairing.
Since  nematicity appears as the composite order parameter of Eq.~\eqref{Eq:nematic-SC}, we expect that the two transitions can be separated, i.e., the nematic transition can occur at a higher temperature than the superconducting transition, giving rise to a vestigial nonsuperconducting phase with broken rotation symmetry~\cite{Schmalian2018,Fernandes2018,FernandesOrthSchmalian2018}. To assess the possibility of a vestigial nematic phase one must go beyond the theory developed here and consider the fluctuations of superconducting order parameter. We leave this as a direction for the future.

Finally, we notice that, alternatively to the strongly-correlated scenario considered in this paper, NSC may, in principle, originate from the internal strain. Indeed, symmetry allows the coupling between the nematic subsidiary order~\eqref{Eq:nematic-SC} and the components of the strain tensor $u_{ij}$ of the form~\cite{VenderbosHc22016}

\be
F_{\text{strain}} = g[(u_{xx}^2-u_{yy}^2)N_1 +2u_{xy}N_2].
\ee
It is clear from this expression that in the presence of uniaxial strain, $u_{xx}^2 - u_{yy}^2 \ne 0,$ one of the superconducting components develops order at higher temperature than the other one, thus resulting in nematic superconductivity. If strain is not too strong, the effect of nematicity becomes weaker as one lowers the temperature~\cite{VenderbosHc22016}, and eventually NSC transits into a chiral superconductor at sufficiently small temperature. To distinguish between the scenarios for NSC from density wave fluctuations and from internal strain, as well as from the strain-induced nematicity in $s$-wave superconductor, a detailed study of the upper critical field behavior in different regimes is required.

%
%

\section{Acknowledgments}
We greatly acknowledge the discussions with Pablo Jarillo-Herrero, Yuan Cao, Daniel Rodan-Legrain, and Oriol Rubies-Bigorda, who shared with us their unpublished experimental data. We also thank Rafael Fernandes, Jonathan Ruhman, and Max Metlitski for numerous valuable discussions. This work is supported by DOE Office of Basic Energy Sciences, Division of Materials Sciences and Engineering under Award DE-SC0018945. LF is partly supported by the David and Lucile Packard Foundation. J.W.F.V. was supported by the National Science Foundation MRSEC
Program, under Grant No.~DMR-1720530.

\appendix

\begin{widetext}

\section{Nematic $d$-wave superconductivity in presence of  charge density wave fluctuations \label{SMSec:CDW}}

In this appendix, we present a detailed calculation of the feedback correction to the free energy of a two-component superconductor due to the presence of charge density wave (CDW) fluctuations. For definiteness, we focus on a $d$-wave superconductor. First, we derive the effective imaginary time action that describes the interplay between pairing potential and the density wave fluctuations. Next, we calculate the very general expression for the feedback contribution to free energy. Finally, we apply our results to particular models which are relevant to twisted bilayer graphene.

\subsection{The model description and the effective action}

As was discussed in the main text, the presence of density wave fluctuations allows us to focus on the vicinities of the points on the Fermi surface connected by the density wave ordering wave vectors, so-called hot spots. The corresponding regions in momentum space near hot spots are called patches. We consider a 2D model with six nonequivalent hot spots in the Brillouin zone, with the CDW wavevectors connecting adjacent hot spots, see Fig.~\ref{SMFig:hotspots}. Then, Hamiltonian~(\ref{Eq:H-A})-(\ref{Eq:H-B}) of the main text in the low-energy limit translates into the imaginary time action for six patches

\be
S = S_{\psi} + S_{\phi}  + S_{\psi - \Delta} + S_{\psi-\phi} . \label{SMeq:S}
\ee
The first term describes the noninteracting electrons near hot spots:
\be
S_\psi = T \sum_{i=1}^6 \sum_{\omega_n, \bk} \left( -i \omega_n + \xi_{i \bk} \right) \psi^\dagger_{i \alpha}(\omega_n, \bk) \psi_{i\alpha}(\omega_n, \bk).
\ee
Here, $\omega_n = 2\pi T (n+1/2)$ are fermionic Matsubara frequencies, index $i = 1,..,6$ numerates hot spots, $\xi_i(\bk)$ is a dispersion near the $i$-th hot spot, and the summation over repeated spin indices $\alpha = \uparrow,\downarrow$ is implied. We assume the presence of sixfold rotational symmetry in the system, which implies that the dispersions near adjacent hot spots are related as
$\xi_i(\bk) = \xi_{i+1}(R_6 \bk)$, where $R_6 = \{ \{ 1/2, -\sqrt{3}/2\},\{\sqrt{3}/2,1/2\}\}$ is the $\pi/3$ rotation matrix. This results in an additional relation that we will actively use, $\xi_i(-\bk) = \xi_{i+3}(\bk),$ where the hot spot index $i$ is defined mod 6. Finally, we assume that different hot spots with indices $i$ and $j$ are related by a mirror symmetry $M_{ij}$, which leads to another useful equality $\xi_i(\bk) = \xi_j(M_{ij}\bk)$.

The second term in Eq.~(\ref{SMeq:S}) is a quadratic action for CDW fluctuations:
\be
S_\phi = \frac T2 \sum_{i=1}^6 \sum_{\Omega_m, \bk} V_{0i}^{-1}(\bq) \phi_i(\Omega_m,\bq) \phi_i^*(\Omega_m,\bq), \label{SMeq:Sphi}
\ee
where $\Omega_m = 2\pi T m$ are bosonic Matsubara frequencies and index $i=1,..,6$ numerates CDW fluctuations with different ordering wavevectors. The fields $\phi_i(\bq)$ should be viewed as 'shifted' with respect to the global CDW field $\phi(\bq)$ introduced in Eqs.~(\ref{Eq:H-A})-(\ref{Eq:H-B}) of the main text, i.e., $\phi_i(\bq) \equiv \phi (\bQ_i + \bq),$ where CDW wavevectors $\bQ_i$ connect hot spots with indices $i$ and $i+1$ as shown in Fig.~\ref{SMFig:hotspots}(a). We assume that the relevant bosonic momenta are much smaller than the distance between the hot spots, $q \ll Q$. Even though the propagator for the global field $\phi(\bq)$, $\tilde V_0(\bq)$, is peaked at $\bQ_i$, the propagators of fields $\phi_i(\bq)$ are apparently peaked at $\bq = 0$, so $V_{0i}(\bq)$ can be thought of as, e.g.,  Lorentzians with the maximum at $q=0$. The particular form of $V_{0i}(\bq)$, however, is not important for us at this moment. Finally, since the global field $\phi(\br)$ is real, the  different 'shifted' components $\phi_i(\Omega, \bq)$ are not independent, but related according to $\phi_{i+3}(-\Omega,-\bq) = \left[\phi_i(\Omega,\bq)\right]^*$.

The third term in Eq.~(\ref{SMeq:S}) describes the two-component $d$-wave pairing,

\be
S_{\psi-\Delta} = \frac T2\sum_{\omega_n, \bk} \tilde \Delta_{ij} \ve_{\alpha \beta} \psi_{i \alpha}^\dagger(\omega_n, \bk) \psi_{j \beta}^\dagger(-\omega_n, -\bk) + \text{H. c.},
\ee
where $\ve_{\alpha \beta}$ is the Levi-Civita tensor in spin space, and the pairing matrix $\tilde \Delta$ is given by

\be
\tilde \Delta = \Delta \cdot I_{\text{spin}} \otimes \left[ \frac{\sqrt{3}}2 d_1 \left( \begin{array}{cccccc} 0 & 0 & 0 & 1 & 0 & 0 \\ 0 & 0 & 0 & 0 & -1 & 0 \\ 0 & 0 & 0 & 0 & 0 & 0 \\ 1 & 0 & 0 & 0 & 0 & 0 \\ 0 & -1 & 0 & 0 & 0 & 0 \\ 0 & 0 & 0 & 0 & 0 & 0 \end{array}  \right) + d_2 \left( \begin{array}{cccccc} 0 & 0 & 0 & -1/2 & 0 & 0 \\ 0 & 0 & 0 & 0 & -1/2 & 0 \\ 0 & 0 & 0 & 0 & 0 & 1 \\ -1/2 & 0 & 0 & 0 & 0 & 0 \\ 0 & -1/2 & 0 & 0 & 0 & 0 \\ 0 & 0 & 1 & 0 & 0 & 0 \end{array}  \right)  \right]_{\text{patches}}. \label{SMeq:Delta0}
\ee
Here $(d_1, d_2)$ plays the role of a two-component superconducting order parameter, and, again, summation over repeated spin indices $\alpha, \beta = \uparrow, \downarrow$ and patch indices $i,j = 1,..,6$ is implied.

Finally, the coupling between electrons and CDW fluctuations is described by the last term in Eq.~(\ref{SMeq:S}):

\begin{align}
S_{\psi-\phi} = \lambda T^2 &\sum_{i=1}^6 \sum_{\substack{\omega_n, \Omega_m \\ \bk, \bq}} \psi_{i+1 \alpha}^\dagger ( \omega_n + \Omega_m, \bk + \bq) \psi_{i \alpha}(\omega_n, \bk) \phi_i(\Omega_m, \bq) + \text{H. c.} = \nonumber \\ = & T \sum_{i,j=1}^6 \sum_{\substack{\omega_n, \Omega_m \\ \bk, \bq}} \psi_{i \alpha}^\dagger( \omega_n + \Omega_m, \bk + \bq) \psi_{j \alpha}(\omega_n, \bk) \hat \Sigma_{ij}(\Omega_m, \bq), \label{SMeq:Spsiphi}
\end{align}
with $\hat \Sigma(\bq,\Omega)$ defined as

\be
\hat \Sigma (\Omega, \bq) = \lambda T \cdot I_{\text{spin}} \otimes\left( \begin{array}{cccccc} 0 & \phi_4(\Omega, \bq) & 0 & 0 & 0 & \phi_6(\Omega, \bq) \\ \phi_1(\Omega, \bq) & 0 & \phi_5(\Omega, \bq) & 0 & 0 & 0 \\ 0 & \phi_2(\Omega, \bq) & 0 & \phi_6(\Omega, \bq) & 0 & 0 \\ 0 & 0 & \phi_3(\Omega, \bq) & 0 & \phi_1(\Omega, \bq) & 0 \\ 0 & 0 & 0 & \phi_4(\Omega, \bq) & 0 & \phi_2(\Omega, \bq) \\ \phi_3(\Omega, \bq) & 0 & 0 & 0 & \phi_5(\Omega, \bq) & 0 \end{array}  \right)_{\text{patches}}. \label{SMeq:hatSigma}
\ee

\begin{figure}
\centerline{\includegraphics[width=.9\textwidth]{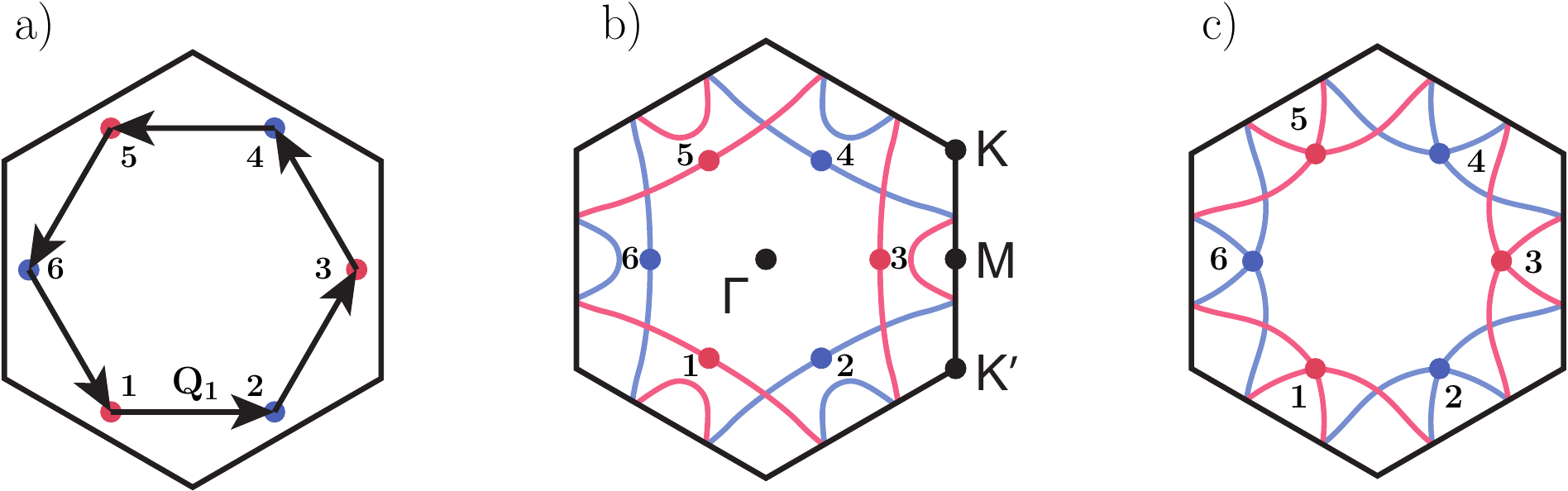}}
\caption{(a) Six inequivalent hot spots in the Brillouin zone are connected by CDW wavevectors $\bQ_i$. All $\bQ_i$ connect hot spots with numbers $i$ and $i+1$, and can be obtained from $\bQ_1$ by sixfold rotations.   (b) Fermi surface in twisted bilayer graphene above the Van Hove singularity. (c) Fermi surface of twisted bilayer graphene at Van Hove energy. This  scenario is realized when the filling factor is close to $n=2$ electrons/holes per supercell. Blue and red parts of the Fermi surface originate from different valleys.}
\label{SMFig:hotspots}
\end{figure}

To proceed, we introduce Nambu particle-hole space according to
\be
\Psi_{i}(\omega, \bk) = \left( \begin{array}{c} \psi_{i \uparrow}(\omega, \bk) \\ \psi_{i\downarrow}(\omega, \bk)  \\ \psi_{i+3\downarrow}^\dagger(-\omega, -\bk) \\   -\psi_{i+3\uparrow}^\dagger(-\omega, -\bk) \end{array}   \right)_N. \label{SMeq:Nambu}
\ee
Then, using the identity $\xi_i(-\bk) = \xi_{i+3}(\bk)$, the action~(\ref{SMeq:S}) can be conveniently rewritten as

\begin{align}
S_{\psi} = -&\frac T2\sum_{i=1}^6 \sum_{\omega_n, \bk} \Psi_{i \alpha}^\dagger(\omega_n, \bk) G^{-1}_{0i}(\omega_n, \bk) \Psi_{i \alpha}(\omega_n, \bk), \nonumber \\
S_{\psi-\Delta} = &\frac T2  \sum_{\omega_n, \bk}\Psi^\dagger(\omega_n, \bk) \Sigma_\Delta \Psi( \omega_n, \bk), \nonumber \\
S_{\psi-\phi} = &\frac T2 \sum_{\substack{\omega_n, \Omega_m \\ \bk, \bq}} \Psi^\dagger(\omega_n + \Omega_m, \bk + \bq) \Sigma_{\phi}(\Omega_m, \bq) \Psi(\omega_n, \bk)   , \label{SMeq:Spsiall}
\end{align}
with

\be
G_{0i}(\omega, \bk) = -\frac{i\omega + \xi_{i\bk} \tau_z}{\omega^2 + \xi_{i\bk}^2}\otimes I_{\text{spin}}, \qquad \Sigma_\Delta = \left( \begin{array}{cc} 0 & \hat \Delta \\ \hat \Delta^\dagger & 0  \end{array}  \right)_N, \qquad \Sigma_\phi(\Omega, \bq) = \hat \Sigma(\Omega, \bq) \otimes \tau_z.  \label{SMeq:Nambuselfenergies}
\ee
Here, $\hat \Sigma$ is given by Eq.~(\ref{SMeq:hatSigma}), $\tau_i$ are Pauli matrices in Nambu space, and $\hat \Delta$ equals

\be
\hat \Delta = \Delta \cdot I_{\text{spin}} \otimes \left[ \frac{\sqrt{3}}2 d_1 \left( \begin{array}{cccccc} 1 & 0 & 0 & 0 & 0 & 0 \\ 0 & -1 & 0 & 0 & 0 & 0 \\ 0 & 0 & 0 & 0 & 0 & 0 \\ 0 & 0 & 0 & 1 & 0 & 0 \\ 0 & 0 & 0 & 0 & -1 & 0 \\ 0 & 0 & 0 & 0 & 0 & 0 \end{array}  \right) + d_2 \left( \begin{array}{cccccc} -1/2 & 0 & 0 & 0 & 0 & 0 \\ 0 & -1/2 & 0 & 0 & 0 & 0 \\ 0 & 0 & 1 & 0 & 0 & 0 \\ 0 & 0 & 0 & -1/2 & 0 & 0 \\ 0 & 0 & 0 & 0 & -1/2 & 0 \\ 0 & 0 & 0 & 0 & 0 & 1 \end{array}  \right)  \right]_{\text{patches}}. \label{SMeq:Delta}
\ee
We note that $\hat \Delta$ has matrix structure different from $\tilde \Delta$, Eq.~(\ref{SMeq:Delta0}), because Nambu space introduced in Eq.~(\ref{SMeq:Nambu}) mixes different patch indices.

Combining Eqs.~(\ref{SMeq:Sphi}) and (\ref{SMeq:Spsiall}), action~(\ref{SMeq:S}) takes simple form

\be
S = S_{\phi} + \frac T2 \text{Tr} \, \Psi^\dagger \left(  -G_0^{-1} +  \Sigma_{\Delta} + \Sigma_{\phi} \right) \Psi,
\ee
where $G_0 = \text{diag}\{ G_{0i} \}_{\text{patches}}$, and $\Tr$ implies trace over spin, Nambu, and patch indices, as well as summation over momenta and frequencies.

To derive the effective theory that describes interplay between CDW fluctuations $\phi_i$ and SC order parameter $d_i$, we integrate out fermions:

\begin{align}
\int D\psi^\dagger D\psi \exp \left[-\frac T2 \Psi^\dagger \left(  -G_0^{-1} +  \Sigma_{\Delta} + \Sigma_{\phi} \right) \Psi \right] = \left\{\text{Det}\left[ T \left(  -G_0^{-1} +  \Sigma_{\Delta} +  \Sigma_{\phi} \right)  \right]\right\}^{1/2} =  \nonumber \\ = \exp\left\{ \frac12 \text{Tr} \ln  \left[ T \left(  -G_0^{-1} +  \Sigma_{\Delta} +  \Sigma_{\phi} \right) \right] \right\} = \left[\text{Det}\left( - T G_0^{-1} \right) \right]^{1/2} \exp \left[\frac12 \text{Tr} \ln  \left(  1 - G_0  \Sigma_{\Delta} -G_0  \Sigma_{\phi} \right) \right].
\end{align}
Neglecting the normal-state electronic part of the partition function, $\left[\text{Det}\left( - T G_0^{-1} \right)\right]^{1/2} $ (which does not depend on $\phi$ or $\Delta$), we find that CDW fluctuations in the presence of pairing potential are described by the effective partition function $Z_{\text{eff}}$ given by

\be
Z_{\text{eff}} = \int D \phi \exp \left[ -S_{\phi} + \frac12\text{Tr} \ln  \left(  1 - G_0  \Sigma_{\Delta} -G_0  \Sigma_{\phi} \right) \right].
\ee
From this expression, we can extract the effective action:

\be
S_{\text{eff}} = S_{\phi} -\frac12 \text{Tr} \ln\left(     1 - G_0  \Sigma_{\Delta} -G_0  \Sigma_{\phi}  \right). \label{SMeq:SphiDeltalog}
\ee

All transformations so far have been exact. Now we make some assumptions that allow us to proceed with our calculation. First, we consider the vicinity of the superconducting transition temperature $T_c$, so we expand the effective action in powers of small pairing potential $\Delta$ (equivalently, in powers of $\Sigma_{\Delta}$) up to fourth order. Second, we assume that CDW fluctuations, though strong, remain massive. Hence, we expand the effective action up to second order in $\phi$ (equivalently, in  $\Sigma_\phi$).

Expanding the logarithm in Eq. (\ref{SMeq:SphiDeltalog}), we find

\begin{align}
S_{\text{eff}} \approx &S_\Delta + S_\phi + \delta S_\phi + S_{\phi - \Delta}, \nonumber \\ \delta S_\phi  = &\frac14 \Tr\left(  G_0 \Sigma_\phi\right)^2, \qquad S_\Delta = \frac14\Tr \left( G_0 \Sigma_\Delta \right)^2 + \frac18 \Tr \left( G_0 \Sigma_\Delta \right)^4, \nonumber \\ S_{\phi - \Delta} = &\frac12 \Tr\left[(G_0 \Sigma_\Delta)^2(G_0 \Sigma_\phi)^2  \right] + \frac14 \Tr [(G_0 \Sigma_\Delta G_0 \Sigma_\phi)^2] + \nonumber \\ + &\frac12 \Tr[(G_0 \Sigma_\phi)^2  (G_0 \Sigma_\Delta)^4] + \frac12 \Tr[(G_0 \Sigma_\Delta)^2 (G_0 \Sigma_\Delta G_0 \Sigma_\phi)^2] + \frac14 \Tr[(G_0 \Sigma_\Delta G_0 \Sigma_\Delta G_0 \Sigma_\phi)^2]. \label{SMeq:SphiDelta}
\end{align}
Here, $S_\Delta$ is a weak-coupling part of Ginzburg-Landau free energy of a superconductor. Terms $S_\phi$ and $\delta S_\phi$ are bare bosonic propagator, Eq.~(\ref{SMeq:Sphi}), and the normal-state CDW susceptibility, respectively. Finally, $S_{\phi - \Delta}$ describes the interplay between CDW fluctuations and superconductivity, which eventually results in the feedback correction to free energy. The different terms in $S_{\phi - \Delta}$ are shown diagrammatically in Fig.~\ref{SMFig:diagrams} and will be explicitly calculated below.

\begin{figure}
\centerline{\includegraphics[width=.8\textwidth]{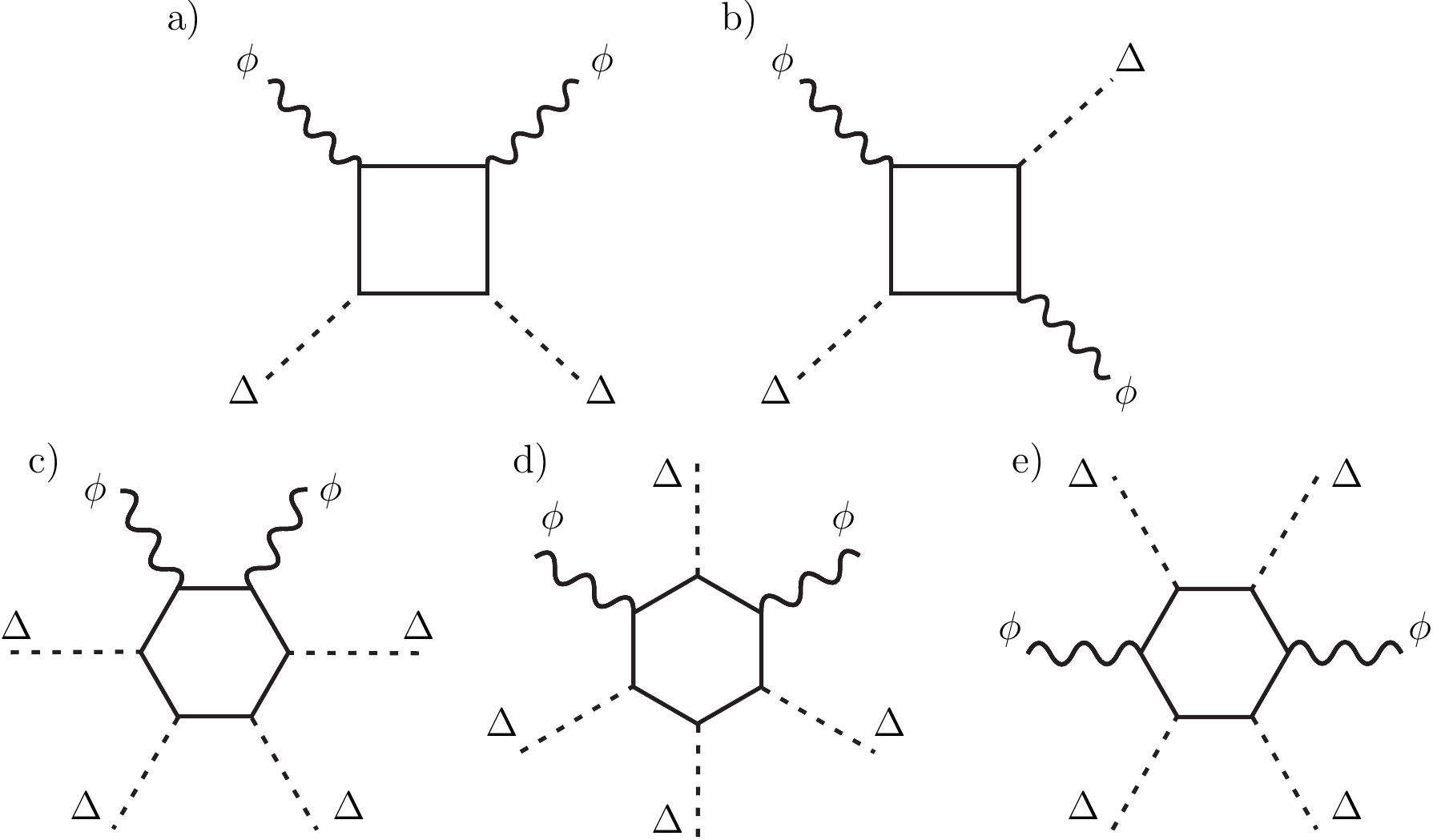}}
\caption{ Diagrams describing the coupling between pairing potential $\Delta(d_1, d_2)$ and density wave fluctuations $\phi_i$. Diagrams (a)-(b) schematically represent the coupling $\phi^2 d^2$ and describe the $\Delta^2$ correction to CDW susceptibility, while diagrams (c)-(e) correspond to the coupling $\phi^2 d^4$ contributing the $\Delta^4$ correction to CDW susceptibility. The contribution to the feedback free energy from  diagrams (a)-(b) becomes dominant  when fluctuations become sufficiently strong.}
\label{SMFig:diagrams}
\end{figure}

\subsection{Weak-coupling analysis \label{Sec:w-c}}

In this section, we calculate the fourth-order weak-coupling contribution to free energy. The corresponding part of the effective action is given by

\be
S^{(4)}_{\Delta} =  \frac18 \Tr[(G_0 \Sigma_{\Delta})^4] = X_0 \Delta^4\left[ 2 \left(|d_1|^2 + |d_2|^2\right)^2 + |d_1^2 + d_2^2|^2 \right], \qquad X_0 = \frac3{8}\sum_{\omega_n, \bk} \left( \frac1{\omega_n^2 + \xi_{i\bk}^2} \right)^2 > 0, \label{SMeq:Swc}
\ee
where $\xi_{i\bk}$ is a dispersion near any of the hot spots, and no summation over $i$ is needed here. The weak-coupling free energy then equals

\be
F^{(4)}_{\Delta} = - T \ln Z^{(4)}_{\Delta} = T X_0 \Delta^4\left[ 2 \left(|d_1|^2 + |d_2|^2\right)^2 + |d_1^2 + d_2^2|^2 \right].
\ee
This result is a 'hot spots' version of Eq.~(\ref{Eq:F0}) of the main text. We see that, since $X_0 > 0$, the weak-coupling approximation favors the fully gapped chiral state, $(d_1, d_2) \sim (1, i)$.

\subsection{ Coupling between pairing potential and CDW fluctuations}

In this section, we present the general expressions for the terms that describe the interplay between CDW fluctuations $\phi_i$ and SC order parameter $d_i$, see Eq.~(\ref{SMeq:SphiDelta}). After straightforward calculation, we find:

\begin{subequations}
\begin{align}
&\Tr \left[ (G_0 \Sigma_\Delta)^2 (G_0 \Sigma_\phi)^2 \right] = 8(\lambda T \Delta)^2 \sum_{\Omega, \bq} |\phi_1(\Omega,\bq)|^2  \left[ K_1(\Omega,\bq) \left| \frac{\sqrt{3}}2d_1 - \frac{1}{2}d_2  \right|^2  + K_1(\Omega,-M\bq) \left| \frac{\sqrt{3}}2d_1 + \frac{1}{2} d_2  \right|^2   \right] + \nonumber  \\ & + |\phi_2(\Omega,\bq)|^2 \left[ K_1(\Omega, -R_6 M \bq) \left| d_2 \right|^2  + K_1(\Omega,R_6^{-1}\bq) \left| \frac{\sqrt{3}}2d_1 + \frac{1}{2}d_2  \right|^2   \right] \nonumber + \\ & + |\phi_3(\Omega,\bq)|^2 \left[ K_1(\Omega,-R_6 \bq) \left| d_2 \right|^2  + K_1(\Omega,M R_6\bq) \left| \frac{\sqrt{3}}2 d_1 - \frac{1}{2} d_2  \right|^2   \right] , \label{SMeq:couplingterms1}\\ &\Tr \left[ (G_0 \Sigma_\Delta  G_0 \Sigma_\phi)^2   \right] = -8(\lambda T \Delta)^2 \sum_{\Omega,\bq} |\phi_1(\Omega,\bq)|^2 K_2(\Omega,\bq) \left[ -\frac{3}2|d_1|^2 + \frac{1}2 |d_2|^2 \right] + \nonumber  \\ & + |\phi_2(\Omega,\bq)|^2 K_2(\Omega,R_6^{-1} \bq) \left[ -|d_2|^2 -\frac{\sqrt{3}}{2}(d_1 d_2^* + d_2 d_1^*)\right]  + |\phi_3(\Omega,\bq)|^2 K_2 (\Omega, - R_6 \bq) \left[ -|d_2|^2 + \frac{\sqrt{3}}{2 }(d_1 d_2^* + d_2 d_1^*)\right] , \label{SMeq:couplingterms2} \\
&\Tr \left[ (G_0 \Sigma_\Delta)^4 (G_0 \Sigma_\phi)^2 \right] = - 8(\lambda T \Delta^2)^2 \sum_{\Omega, \bq} |\phi_1(\Omega,\bq)|^2 \left[ K_3(\Omega,\bq) \left| \frac{\sqrt{3}}2 d_1 - \frac{1}{2} d_2  \right|^4  + K_3(\Omega, -M\bq) \left| \frac{\sqrt{3}}2 d_1 + \frac{1}{2} d_2  \right|^4   \right] + \nonumber  \\ & + |\phi_2(\Omega,\bq)|^2 \left[ K_3(\Omega,-R_6 M \bq) \left| d_2\right|^4  + K_3(\Omega,R_6^{-1}\bq) \left| \frac{\sqrt{3}}2 d_1 + \frac{1}{2} d_2  \right|^4   \right] \nonumber + \\ & + |\phi_3(\Omega,\bq)|^2 \left[ K_3(\Omega,-R_6 \bq) \left| d_2 \right|^4  + K_3(\Omega,M R_6\bq) \left| \frac{\sqrt{3}}2 d_1 - \frac{1}{2} d_2  \right|^4  \right], \label{SMeq:couplingterms3} \\ &\Tr \left[(G_0 \Sigma_\Delta)^2 (G_0 \Sigma_\Delta  G_0 \Sigma_\phi)^2   \right] = 4(\lambda T \Delta^2)^2 \sum_{\Omega,\bq} |\phi_1(\Omega,\bq)|^2 \left( -\frac{3}2|d_1|^2 + \frac{1}2|d_2|^2 \right) \times \nonumber \\ \times &\left[  K_4(\Omega,\bq) \left| \frac{\sqrt{3}}2d_1 - \frac{1}{2} d_2  \right|^2+ K_4(\Omega, -M\bq) \left| \frac{\sqrt{3}}2d_1 + \frac{1}{2}d_2  \right|^2 \right] + \nonumber  \\ & + |\phi_2(\Omega,\bq)|^2 \left( -|d_2|^2 -\frac{\sqrt{3}}2(d_1 d_2^* + d_2 d_1^*)\right)\left[ K_4(\Omega, -R_6 M \bq) \left| d_2  \right|^2 + K_4(\Omega, R_6^{-1} \bq) \left| \frac{\sqrt{3}}2d_1 + \frac{1}{2}d_2 \right|^2\right]  + \nonumber  \\ &+ |\phi_3(\Omega,\bq)|^2  \left( -|d_2|^2 +\frac{\sqrt{3}}{2}(d_1 d_2^* + d_2 d_1^*)\right)\left[ K_4(\Omega, - R_6  \bq) \left| d_2  \right|^2 + K_4(\Omega, M R_6 \bq) \left| \frac{\sqrt{3}}2d_1 - \frac{1}{2}d_2  \right|^2\right], \label{SMeq:couplingterms4} \\  &\Tr \left[ (G_0 \Sigma_\Delta G_0 \Sigma_\Delta  G_0 \Sigma_\phi)^2   \right] = -16(\lambda T \Delta^2)^2 \sum_{\Omega,\bq} |\phi_1(\Omega,\bq)|^2 K_5(\Omega,\bq) \left| \frac{\sqrt{3}}2d_1 - \frac{1}2d_2  \right|^2 \left| \frac{\sqrt{3}}2d_1 + \frac{1}2d_2  \right|^2 + \nonumber  \\ & + |\phi_2(\Omega,\bq)|^2  K_5(\Omega,R_6^{-1} \bq) \left| \frac{\sqrt{3}}2d_1 + \frac{1}2d_2 \right|^2 \left| d_2  \right|^2  + |\phi_3(\Omega,\bq)|^2  K_5 (\Omega, -R_6 \bq) \left| \frac{\sqrt{3}}2d_1 - \frac{1}2d_2 \right|^2 \left|  d_2  \right|^2, \label{SMeq:couplingterms5}
\end{align}
\end{subequations}
where, again, $R_6$ is a six-fold rotation matrix, $M \equiv M_{12}$ is a mirror symmetry operation between hot spots 1 and 2, and we used the equalities $\xi_i(\bk) = \xi_{i+1}(R_6 \bk)$ and $\xi_1(\bk) = \xi_2 (M \bk)$ in our derivation. Functions $K_1(\Omega, \bq),..,K_5(\Omega,\bq)$ are defined as

\begin{align}
K_1(\Omega,\bq) \equiv \sum_{\bk, \omega} \frac{\omega (\omega + \Omega) - \xi_{1\bk} \xi_{2 \bk - \bq}}{ \left[\omega^2 + \xi_{1\bk}^2\right]^2\left[ (\omega+ \Omega)^2 + \xi_{2 \bk - \bq}^2 \right]}, \nonumber \\
K_2(\Omega, \bq) \equiv \sum_{\bk, \omega} \frac{1}{ \left[\omega^2 + \xi_{1\bk}^2\right]\left[ (\omega+ \Omega)^2 + \xi_{2 \bk - \bq}^2 \right]}, \nonumber \\
K_3(\Omega,\bq) \equiv \sum_{\bk, \omega} \frac{\omega (\omega + \Omega) - \xi_{1\bk} \xi_{2 \bk - \bq}}{ \left[\omega^2 + \xi_{1\bk}^2\right]^3\left[ (\omega+ \Omega)^2 + \xi_{2 \bk - \bq}^2 \right]}, \nonumber \\
K_4(\Omega, \bq) \equiv \sum_{\bk, \omega} \frac{1}{ \left[\omega^2 + \xi_{1\bk}^2\right]^2\left[ (\omega+ \Omega)^2 + \xi_{2 \bk - \bq}^2 \right]}, \nonumber \\
K_5(\Omega,\bq) \equiv \sum_{\bk, \omega} \frac{\omega (\omega + \Omega) - \xi_{1\bk} \xi_{2 \bk - \bq}}{ \left[\omega^2 + \xi_{1\bk}^2\right]^2\left[ (\omega+ \Omega)^2 + \xi_{2 \bk - \bq}^2 \right]^2}. \label{SMeq:K1-K5}
\end{align}
In our derivation, we also exploited the equalities $K_2(\Omega, -\bq) = K_2(\Omega, M\bq)$ and $K_5(\Omega, -\bq) = K_5(\Omega, M\bq)$.

The first two terms,(\ref{SMeq:couplingterms1})-(\ref{SMeq:couplingterms2}), are given by diagrams in Figs.~\ref{SMFig:diagrams}(a) and \ref{SMFig:diagrams}(b)  and describe the $\Delta^2$ correction to the CDW susceptibility. Terms (\ref{SMeq:couplingterms3})-(\ref{SMeq:couplingterms5}) correspond to the $\Delta^4$ correction to the CDW susceptibility, as represented by diagrams in Figs.~\ref{SMFig:diagrams}(c)-\ref{SMFig:diagrams}(e). We also note that all terms are expressed through fields $\phi_1-\phi_3$ only, since $\phi_4-\phi_6$ can be eliminated using equality $\phi_{i+3}(-\Omega, -\bq) = \left[ \phi_i(\Omega,\bq)  \right]^*$.

Though functions $K_1 - K_5$ depend on the dispersion relations near hot spots, $\xi_{i\bk}$, which we have not specified yet, a very general conclusion regarding the favorable superconducting state can be drawn without an explicit evaluation of $K_i$.

\subsection{ Feedback correction to free energy}

Total free energy of CDW fluctuations in the presence of pairing is given by

\be
F_{\text{CDW}} = -T \ln Z_{\text{CDW}}, \qquad Z_{\text{CDW}} \equiv \int D\phi \exp[-(S_{\text{eff}} - S_{\Delta})],
\ee
where we subtracted the weak-coupling contribution~(\ref{SMeq:Swc}).

Assuming that CDW fluctuations remain massive, we explicitly rewrite the effective action (\ref{SMeq:SphiDelta}) as a quadratic form of $\phi_i$,
\be
S_{\text{eff}} - S_{\Delta} \approx T \sum_{i,j=1}^3 \sum_{\Omega,\bq} \phi_i(\Omega,\bq)\left\{ \left[ \hat V^{-1}(\Omega, \bq) \right]^{ij} - \delta\hat \chi^{ij}(\Omega,\bq) \right\} \phi_j^*(\Omega,\bq),
\ee
where $\hat V(\Omega, \bq)$ is the normal-state bosonic propagator which includes the normal-state polarization operator, and $\delta \hat \chi (\Omega, \bq)$ is a superconducting correction to the CDW susceptibility determined by Eqs.~(\ref{SMeq:couplingterms1})-(\ref{SMeq:couplingterms5}). Again, we expressed $S_{\text{eff}}$ in terms of three independent complex fields $\phi_1(\Omega, \bq) - \phi_3(\Omega, \bq)$ only.

Integrating out fields $\phi_i$ and expanding the result in powers of $\delta \hat \chi$, or, equivalently, in powers of $\Delta$, one easily obtains

\be
F_{\text{CDW}} = F^{(0)} + \delta F^{(2)}_{\Delta} + \delta F^{(4)}_{\Delta} + \ldots = F^{(0)} - T  \Tr \left( \hat V \delta \hat \chi \right) - \frac{T}2 \Tr( \hat V \delta \hat \chi)^2 + \ldots, \label{SMeq:Fphi}
\ee
where $\ldots$ stands for the terms of order $O(\Delta^6)$.

The first term in Eq. (\ref{SMeq:Fphi}), $F^{(0)}$, does not depend on pairing potential and gives the energy of CDW fluctuations in the normal state. To evaluate second and third terms, we assume that the bosonic propagator is diagonal in patch space. Then, due to rotational symmetry, it has the form

\be
\hat V(\Omega, \bq) = \left( \begin{array}{ccc} V(\Omega, \bq)& 0 & 0 \\ 0& V(\Omega, R_6^{-1}\bq) & 0 \\ 0 & 0 & V(\Omega, R_6^{-2}\bq) \end{array} \right)_{\text{patches}}.
\ee
Assuming further that $V(\Omega, \bq)$ satisfies $V(\Omega, -M \bq) = V(\Omega, \bq)$, we find for $\delta F_{\Delta}^{(2)}$

\begin{align}
 \delta F_{\Delta}^{(2)} &\equiv -T \Tr\left( \hat V \delta \hat \chi \right) =  3  (\lambda T \Delta)^2 (4 X_1 + X_2)(|d_1|^2 + |d_2|^2) - \nonumber \\ &- \frac{3  (\lambda T \Delta^2)^2}4\left[ (8 X_3 + 4 X_4 + 4 X_5)(|d_1|^2 + |d_2|^2)^2 + \left( 4 X_3 + 2 X_4 - X_5 \right)|d_1^2+d_2^2|^2  \right], \label{SMeq:F^2}
\end{align}
where we defined $X_i \equiv \sum_{\Omega, \bq} V(\Omega, \bq) K_i(\Omega, \bq)$.

Analogously, using the equality $K_2(\Omega, -\bq) = K_2(\Omega, M\bq)$, we find the expression  for $ \delta F_{\Delta}^{(4)}$

\be
\delta F_{\Delta}^{(4)} \equiv - \frac{T}2 \Tr( \hat V \delta \hat \chi)^2 = -\frac{3T^3 (\lambda  \Delta)^4}2\left[\left(|d_1|^2 + |d_2|^2  \right)^2 (Y_1 + 8 Y_2 + 8 Y_3 + 8 Y_4) + \left|d_1^2 + d_2^2  \right|^2 (2 Y_1 + 4 Y_2 + 4 Y_3 - 2 Y_4)  \right], \label{SMeq:F^4}
\ee
with

\begin{align}
 &Y_1 \equiv \sum_{\bq, \Omega} V^2(\Omega,\bq) K_2^2(\Omega, \bq), \qquad &&Y_2 \equiv \sum_{\bq, \Omega} V^2(\Omega,\bq) K_1^2(\Omega, \bq), \nonumber \\ &Y_3 \equiv \sum_{\bq, \Omega} V^2(\Omega,\bq) K_1(\Omega, \bq) K_2(\Omega, \bq), \qquad &&Y_4 \equiv \sum_{\bq, \Omega} V^2(\Omega,\bq) K_1(\Omega, \bq) K_1(\Omega, - M \bq). \label{SMeq:Y1-Y4}
\end{align}

It is straightforward to show that $Y_1 + Y_2 \geq 2 |Y_3|$ and $Y_2 \geq |Y_4|,$ which leads to $Y_1 + 2 Y_2 + 2 Y_3 - Y_4 \geq 0$. Hence, the correction to free energy $\delta F_{\Delta}^{(4)}$ {\it always} favors a nematic superconducting state. As we demonstrate below using specific examples, the correction $\delta F_{\Delta}^{(2)}$ becomes parametrically smaller than $\delta F_{\Delta}^{(4)}$ once the fluctuations become sufficiently strong.

If $V(\Omega, \bq)$ is strongly peaked at $\Omega = q =0$,  only the zeroth mode significantly contributes to free energy. In this case, the coupling between CDW fluctuations and SC order parameter is given by (we neglect $\phi^2 d^4$ terms for now)

\begin{align}
S_{\phi-\Delta} & \approx \frac12 \Tr\left[(G_0 \Sigma_\Delta)^2(G_0 \Sigma_\phi)^2  \right] + \frac14 \Tr [(G_0 \Sigma_\Delta G_0 \Sigma_\phi)^2] = \nonumber \\ & = (\lambda T \Delta)^2 \left\{ |\boldsymbol{\phi}|^2 |\bd|^2 [4 K_1(0,0) + K_2(0,0)] + [P_1 N_1 + P_2 N_2] [K_1(0,0) + K_2(0,0)]\right\}, \label{SMeq:q=0 coupling}
\end{align}
in agreement with Eqs.~(\ref{Eq:deltaFphiDelta}) and (\ref{Eq:beta12}) of the main text. Here we defined

\begin{align}
&|\boldsymbol{\phi}|^2 \equiv |\phi_1(0,0)|^2 + |\phi_2(0,0)|^2 + |\phi_3(0,0)|^2, \quad &&|\bd|^2 \equiv |d_1|^2 + |d_2|^2, \nonumber \\
& N_1 \equiv |d_1|^2 - |d_2|^2, \qquad && N_2 \equiv d_1 d_2^* + d_2 d_1^*, \nonumber \\
& P_1 \equiv 2 |\phi_1(0,0)|^2  - |\phi_2(0,0)|^2 - |\phi_3(0,0)|^2 , \qquad && P_2 \equiv \sqrt{3} \left(  |\phi_2(0,0)|^2 - |\phi_3(0,0)|^2\right).
\end{align}

Integrating out $\phi_i$, we obtain

\be
\delta F_{\Delta}^{(4)} \approx -\frac{3T^3 (\lambda \Delta)^4}2 \left[ \int \frac{d^2q}{(2\pi)^2} V^2(0,q) \right] \left\{ \left[ 4 K_1(0,0)+K_2(0,0)  \right]^2 (|d_1|^2+|d_2|^2)^2 + 2 \left[ K_1(0,0) + K_2(0,0) \right]^2|d_1^2 + d_2^2|^2\right\},
\ee
in agreement with Eq.~(\ref{Eq:FphiDeltaaveraged}) of the main text. We see that the coupling between nematic bilinears $N_i P_i$ is an essential ingredient that eventually leads to nematic superconductivity. As we demonstrate below within two explicit models, which we believe adequately describe the low-energy physics in twisted bilayer graphene at different dopings and twist angles, the quartic ($\sim \Delta^4$) term from $\delta F_{\Delta}^{(2)}$ (originating from the coupling $\phi^2 d^4$) becomes negligible compared to $\delta F_{\Delta}^{(4)}$ as the fluctuation strength increases.

\subsection{Application for specific model: hot spots with linear dispersion \label{SMSec:specificmode1}}

Now we explicitly calculate the feedback free energy for the model of twisted bilayer graphene with Fermi surface slightly away from the Van Hove singularity. In this case, the Fermi surface is  shown in  Fig.~\ref{SMFig:hotspots}(b), with the single-electron dispersion near hot spots approximated by $\xi_i(\bk) = v (\bk \cdot \bn_i)$, where $\bn_i$ is a unit vector in the $\Gamma\text{M}_i$ direction. For concreteness, we choose the coordinates such that

\be
\xi_1(\bk) = -\frac{v}2\left( k_x + \sqrt{3} k_y    \right), \qquad \xi_2(\bk) = -\frac{v}2\left( -k_x + \sqrt{3} k_y    \right).
\ee

With these explicit expressions for $\xi_{i\bk}$, we can calculate $K_i(\Omega, \bq)$ defined in Eq.~(\ref{SMeq:K1-K5}). Integrals over $\bk$ can be readily evaluated, giving

\begin{align}
&K_1(\Omega_m,\bq) = \frac1{4\sqrt{3} v^2}\frac1{(2\pi T)^2}\sum_n\frac{\text{sign}(n+1/2) \text{sign}(n+m+1/2)}{(n+1/2)^2} = \frac1{2\sqrt{3} v^2}\frac1{(2\pi T)^2} \psi_0'\left(|m|+\frac12  \right), \nonumber \\
&K_2(\Omega_m, \bq) = \frac1{2\sqrt{3} v^2}\frac1{(2\pi T)^2}\sum_n\frac{1}{|n+1/2||n+m+1/2|} = \frac1{2\sqrt{3} v^2}\frac1{(2\pi T)^2} \left\{ \begin{array}{cc} 4\left[ \psi_0\left( |m|+\frac12 \right) - \psi_0 \left( \frac12 \right) \right]/|m|, & m \ne 0 \\ \pi^2, & m=0  \end{array}  \right., \nonumber \\
&K_3(\Omega_m,\bq) = \frac{\sqrt{3}}{16 v^2}\frac1{(2\pi T)^4}\sum_n\frac{\text{sign}(n+1/2) \text{sign}(n+m+1/2)}{(n+1/2)^4}, \nonumber \\
&K_4(\Omega_m, \bq) = \frac1{4\sqrt{3} v^2}\frac1{(2\pi T)^4}\sum_n\frac{1}{|n+1/2|^3|n+m+1/2|}, \nonumber \\
&K_5(\Omega_m,\bq) = \frac1{8\sqrt{3} v^2}\frac1{(2\pi T)^4}\sum_n\frac{\text{sign}(n+1/2) \text{sign}(n+m+1/2)}{(n+1/2)^2 (n+m+1/2)^2}.
\end{align}
Here $\psi_0(x)$ is the digamma function, and $\Omega_m = 2\pi T m$. In principle, frequency dependence of $K_3 - K_5$ can also be expressed through $\psi_0$ and its derivatives. The resulting expressions, however, are rather cumbersome, and we do not present them here.

To calculate the explicit expression for the feedback free energy, we further assume that the bosonic propagator is given by

\be
V(\Omega_m, \bq) = \frac{\chi_0}{c(q_0^2 + q^2) + \Omega_m^2}, \qquad \text{with} \qquad   T \ll cq_0 \ll c k_{\max},
\ee
where $k_{\max}$ is an ultraviolet momentum cutoff corresponding to the size of patches.

Corrections $X_1$ and $X_2$ in Eq.~(\ref{SMeq:F^2}) only shift $T_c$, so we do not consider them here. Then, it can be directly shown that the most singular contribution to $\delta F_\Delta^{(2)}$ comes from $X_4$, since $K_4(\Omega_m,\bq) \sim 1/m$ for $m\gg1$. Consequently, omitting terms $\sim \Delta^2$, we find for $\delta F_\Delta^{(2)}$

\be
\delta F_{\Delta}^{(2)} \to -\frac{3 (\lambda T \Delta^2)^2}{2}X_4\left[ 2(|d_1|^2 + |d_2|^2)^2 + |d_1^2 + d_2^2|^2    \right], \qquad X_4 = \frac{7 \zeta(3)}{4\sqrt{3}\pi} \left( \frac1{2\pi T} \right)^4 \frac{\chi_0}{c^2 v^2}\ln \frac{k_{\max}}{q_0} \ln \frac{c^2 k_{\max} q_0}{T^2}.
\ee

Analogously, performing integration over $\bq$ and summation over $\Omega_m$, we find

\be
Y_1 = 3.23 \left(\frac{\chi_0}{q_0 v^2 c^2} \right)^2 \cdot\left(\frac1{2\pi T}  \right)^4, \qquad Y_2 = Y_4 = 0.18 \left(\frac{\chi_0}{q_0 v^2 c^2} \right)^2 \cdot \left(\frac1{2\pi T}  \right)^4, \qquad Y_3 = 0.54\left(\frac{\chi_0}{q_0 v^2 c^2} \right)^2 \cdot \left(\frac1{2\pi T}  \right)^4,
\ee
leading to

\be
\delta F_\Delta^{(4)} = - \frac{13.46}{ (2\pi)^4}\left(\frac{\chi_0 T \lambda^2}{ q_0 v^2 c^2} \right)^2 \frac{\Delta^4}{T^3} \left[ 1.16 (|d_1|^2 + |d_2|^2 )^2 + |d_1^2 + d_2^2|^2  \right].
\ee
This result is presented in Eq.~(\ref{Eq:answer2}) of the main text. We see that $\delta F_\Delta^{(4)}$ becomes dominant over $\delta F_\Delta^{(2)}$ (at fourth-order in $\Delta$) when fluctuations become sufficiently strong, i.e., when $q_0$ becomes sufficiently small. This observation allows us to focus on the $\delta F_\Delta^{(4)}$ term in this paper.

\subsection{Application for  specific model: hot spots at Van Hove singularities \label{SMSec:specificmode2}}

Next, we consider a model of twisted bilayer graphene with the filling factor close to $n=2$ electron/holes per supercell. In this case, hot spots coincide with the Van Hove singularities, and the representative Fermi surface  is shown in Fig.~\ref{SMFig:hotspots}(c). We assume for simplicity that the dispersion near Van Hove points is the same as in the case of monolayer graphene, i.e., $\xi_{1,2}(\bk) = 3t(k_y^2 \pm \sqrt{3}k_x k_y)/2,$ where $t$ is an effective hopping constant, and the hyperlattice constant is absorbed into the definition of $t$.  While we consider sufficiently strong CDW fluctuations, we stay away from the immediate vicinity of the transition into the CDW-ordered state. Hence, we assume that the CDW propagator is given by

\be
V(\Omega, q) = \frac{\chi_0}{c^2( q_0^2 + q^2)}, \qquad  T/t k_{\max} \ll q_0 \ll \sqrt{T/t},
\ee
where, again, $k_{\max}$ is an effective ultraviolet momentum cutoff. The relevant bosonic momenta and frequencies then satisfy $q \lesssim \sqrt{T/t}$ and $\Omega \sim T$, respectively. In this range of frequencies and momenta, the asymptotic behavior of functions $K_i(\Omega, \bq)$ defined in Eq.~(\ref{SMeq:K1-K5}) is given by
\be
K_i(\Omega,\bq)= \frac1{3 \sqrt{3} \pi t} f_i(\Omega) \ln \left( \min \left\{ k_{\max} \sqrt{\frac{t}T}  , \frac1{|q_y|} \sqrt{ \frac{T}t} \right\}\right),  \qquad q\lesssim \sqrt{\frac{T}t}, \quad \Omega \sim T. \label{SMeq:K_i}
\ee
The main contribution to $K_i$ comes from the 'nesting' direction, $k_y \approx 0$. Frequency-dependent functions $f_i(\Omega)$ are given by

\begin{align}
&f_1(\Omega_m) =\frac12 \sum_{\omega_n}\frac{2+2 \sign[ \omega_n (\omega_n + \Omega_m)] + \frac{\Omega_m}{\omega_n} }{|\omega_n|(|\omega_n| + |\omega_n + \Omega_m|)^2} = \frac{1}{(2\pi T)^3} \cdot \left\{\begin{array}{cc} 0, & m\ne 0, \\ 7 \zeta(3), & m = 0  \end{array}  \right. \nonumber\\
&f_2(\Omega_m) = \sum_{\omega_n}  \frac1{|\omega_n||\omega_n + \Omega_m| (|\omega_n| + |\omega_n + \Omega_m|)} = \frac{1}{(2\pi T)^3} \cdot \left\{\begin{array}{cc} \frac4{m^2}\left[\ln 4 + \text{HarmonicNumber}\left( \frac{|m|-1}2  \right)  \right], & m\ne 0, \\ 7 \zeta(3), & m = 0  \end{array}  \right. \nonumber\\
&f_3(\Omega_m) = \frac{1}8\sum_{\omega} \frac{3 \omega_n^2 + |\omega_n (\omega_n + \Omega_m)| + \sign[\omega_n(\omega_n + \Omega_m)]\cdot\left[8\omega_n^2 + 9|\omega_n(\omega_n+\Omega_m)| + 3(\omega_n+\Omega_m)^2  \right]}{\omega_n^4(|\omega_n| + |\omega_n + \Omega_m|)^3} = \nonumber\\
&=\frac{1}{(2\pi T)^5} \left\{\begin{array}{cc} -\frac1{4m^2}\left[\psi_0''\left( \frac{1+|m|}2 \right)+14 \zeta(3) \right], & m\ne 0, \\ 93\zeta(5)/4, & m = 0  \end{array}  \right.   \nonumber \\
&f_4(\Omega_m) = \frac12 \sum_{\omega_n} \frac{2|\omega_n| + |\omega_n + \Omega_m|}{|\omega_n|^3 |\omega_n + \Omega_m| (|\omega_n| + |\omega_n + \Omega_m|)^2} = \nonumber \\ &= \frac{1}{(2\pi T)^5} \cdot \left\{\begin{array}{cc} \frac1{m^4}\left[8\ln 2 + 4\text{HarmonicNumber}\left( \frac{|m|-1}2  \right) + 7 m^2 \zeta(3)\right], & m\ne 0, \\ 93 \zeta(5)/4, & m = 0  \end{array}  \right., \nonumber\\
&f_5(\Omega_m) = \frac12 \sum_{\omega_n} \frac{|\omega_n||\omega_n + \Omega_m| + \sign[\omega_n(\omega_n + \Omega_m)]\cdot\left[\omega_n^2 + 3|\omega_n(\omega_n+\Omega_m)| + (\omega_n+\Omega_m)^2  \right]}{\omega_n^2 (\omega_n + \Omega_m)^2(|\omega_n| + |\omega_n + \Omega_m|)^3} = \nonumber\\
&=\frac{1}{(2\pi T)^5} \left\{\begin{array}{cc} \frac1{2m^4}\left[m^2 \psi_0''\left( 2,\frac{1+|m|}2 \right)-8 \text{HarmonicNumber}\left( \frac{|m|-1}2  \right) - 16 \ln2 \right], & m\ne 0, \\ 93\zeta(5)/4, & m = 0  \end{array}  \right. \label{SMeq:f1-f5}
\end{align}
where, again, $\omega_n = 2\pi T (n+1/2)$, $\Omega_m = 2\pi T m,$ $\zeta(x)$ is the Riemann zeta function, and $\psi_0(x)$ is the digamma function.

After straightforward integration over $\bq$ and summation over $\Omega_m$, we find

\begin{align}
&X_1 = \frac{7\zeta(3)}{2^7 3\sqrt{3} \pi^5}\frac{\chi_0}{t c^2 T^3} \ln^2 \frac{T}{t q_0^2},  &&X_2 =  4 \frac{7\zeta(3)}{2^7 3\sqrt{3} \pi^5}\frac{\chi_0}{t c^2 T^3} \ln^2 \frac{T}{t q_0^2} =  4 X_1, \nonumber \\
&X_3 =  \frac{23.22}{2^{10} 3\sqrt{3} \pi^7}\frac{\chi_0}{t c^2 T^5} \ln^2 \frac{T}{t q_0^2}, &&X_4 = \frac{128.58}{2^{10} 3\sqrt{3} \pi^7}\frac{\chi_0}{t c^2 T^5} \ln^2 \frac{T}{t q_0^2}, \nonumber  \\
&X_5 = \frac{17.85}{2^{10} 3\sqrt{3} \pi^7}\frac{\chi_0}{t c^2 T^5} \ln^2 \frac{T}{t q_0^2}, && \nonumber\\
&Y_1 = \frac{144.57}{2^{10} 3^3 \pi^9} \frac{\chi_0^2}{t^2 c^4 q_0^2 T^6} \ln^2\frac{T}{t q_0^2}, &&Y_2 = Y_3 = Y_4  = \frac{49 \zeta^2(3)}{2^{10} 3^3 \pi^9} \frac{\chi_0^2}{t^2 c^4 q_0^2 T^6} \ln^2\frac{T}{t q_0^2} \approx \frac{70.8}{2^{10} 3^3 \pi^9} \frac{\chi_0^2}{t^2 c^4 q_0^2 T^6} \ln^2\frac{T}{t q_0^2}.
\end{align}

Collecting everything together, we find the feedback correction to free energy:

\be
\delta F_{\Delta}^{(2)} \approx \frac{7 \zeta(3)}{2^4  \sqrt{3} \pi^5}  \frac{\chi_0 \lambda^2 }{t c^2} \frac{\Delta^2}T  \ln^2\frac{T}{t q_0^2}(|d_1|^2 + |d_2|^2)- \frac{6.0}{ (2 \pi)^7}\frac{\chi_0 \lambda^2}{t c^2} \frac{\Delta^4}{T^3} \ln^2 \frac{T}{t q_0^2} \left[ 2.32 (|d_1|^2 + |d_2|^2)^2 + |d_1^2 + d_2^2|^2\right],
\ee

\be
\delta F_{\Delta}^{(4)} \approx - \frac{3.16}{ (2 \pi)^8}\left(  \frac{\chi_0 \lambda^2}{t q_0 c^2} \right)^2 \frac{\Delta^4}{T^3} \ln^2 \frac{T}{t q_0^2} \left[ 2.58 (|d_1|^2 + |d_2|^2)^2 + |d_1^2 + d_2^2|^2\right].
\ee

We see, again, that $\delta F_{\Delta}^{(4)}$ becomes more significant than the fourth-order term in $\delta F_{\Delta}^{(2)}$ as $q_0$ decreases, hence, the latter can be neglected when fluctuations become sufficiently strong.

\section{Nematic $d$-wave superconductivity in presence of  spin density wave fluctuations \label{SMSec:SDW}}

In this appendix, we repeat the analysis of the previous appendix for the case of two-component $d$-wave superconductor coupled to spin density wave fluctuations, instead of charge density wave. Due to spin-rotation invariance, it is sufficient to consider the case of uniaxial SDW only, which we do for simplicity. The answer for the feedback free energy in case of $SU(2)$-symmetric SDW is given by the same expression, up to an overall numerical coefficient.

The whole analysis for the case of SDW fluctuations is very similar to the one presented in the previous section. The only difference is that the coupling between fermions and SDW fluctuations is now given by

\begin{align}
S_{\psi-\phi} = \lambda T^2 &\sum_{\alpha, \beta = \uparrow, \downarrow}\sum_{i=1}^6 \sum_{\substack{\omega_n, \Omega_m \\ \bk, \bq}} \psi_{i+1 \alpha}^\dagger ( \omega_n + \Omega_m, \bk + \bq) \sigma^z_{\alpha \beta} \psi_{i \beta}(\omega_n, \bk) \phi_i(\Omega_m, \bq) + \text{H. c.} = \nonumber \\ = T & \sum_{\alpha, \beta = \uparrow, \downarrow} \sum_{i,j=1}^6 \sum_{\substack{\omega_n, \Omega_m \\ \bk, \bq}} \psi_{i \alpha}^\dagger( \omega_n + \Omega_m, \bk + \bq) \psi_{j \beta}(\omega_n, \bk) \hat \Sigma_{ij, \alpha \beta}(\Omega_m, \bq),
\end{align}
with $\hat \Sigma(\bq,\Omega)$ defined as

\be
\hat \Sigma_{ij, \alpha \beta} (\Omega, \bq) = \lambda T\cdot \sigma^z_{\alpha \beta} \otimes\left( \begin{array}{cccccc} 0 & \phi_4(\Omega, \bq) & 0 & 0 & 0 & \phi_6(\Omega, \bq) \\ \phi_1(\Omega, \bq) & 0 & \phi_5(\Omega, \bq) & 0 & 0 & 0 \\ 0 & \phi_2(\Omega, \bq) & 0 & \phi_6(\Omega, \bq) & 0 & 0 \\ 0 & 0 & \phi_3(\Omega, \bq) & 0 & \phi_1(\Omega, \bq) & 0 \\ 0 & 0 & 0 & \phi_4(\Omega, \bq) & 0 & \phi_2(\Omega, \bq) \\ \phi_3(\Omega, \bq) & 0 & 0 & 0 & \phi_5(\Omega, \bq) & 0 \end{array}  \right)_{ij},
\ee
instead of Eqs.~(\ref{SMeq:Spsiphi})-(\ref{SMeq:hatSigma}), and $\sigma^z$ is a Pauli matrix in spin space. This leads to the self-energy in the Nambu space

\be
\Sigma_\phi(\Omega, \bq) = \hat \Sigma(\Omega, \bq) \otimes I_N,
\ee
instead of the corresponding term in Eq.~(\ref{SMeq:Nambuselfenergies}).

As a consequence of a different structure of $\Sigma_\phi$ in Nambu space, the expressions for $\Tr [(G_0 \Sigma_\Delta G_0 \Sigma_\phi)^2]$ and $\Tr[(G_0 \Sigma_\Delta)^2 (G_0 \Sigma_\Delta G_0 \Sigma_\phi)^2]$ in Eqs.~(\ref{SMeq:couplingterms2}) and (\ref{SMeq:couplingterms4}) contain an overall extra minus sign compared to the case of CDW fluctuations, which can be absorbed by redefining $K_2 \to -K_2$ and $K_4 \to -K_4$. This leads, in particular, to an extra minus sign in front of terms in free energy containing $X_2$, $X_4$, and $Y_3$. More explicitly, the feedback corrections in case of SDW fluctuations are given by

\begin{align}
\delta F_{\Delta}^{(2)} &=  3 (\lambda T \Delta)^2 (4 X_1 - X_2)(|d_1|^2 + |d_2|^2) - \nonumber \\ &- \frac{3  (\lambda T \Delta^2)^2}4\left[ (8 X_3 - 4 X_4 + 4 X_5)(|d_1|^2 + |d_2|^2)^2 + \left( 4 X_3 - 2 X_4 - X_5 \right)|d_1^2+d_2^2|^2  \right],
\end{align}

\be
\delta F_{\Delta}^{(4)} =  -\frac{3T^3 (\lambda  \Delta)^4}2\left[\left(|d_1|^2 + |d_2|^2  \right)^2 (Y_1 + 8 Y_2 - 8 Y_3 + 8 Y_4) + \left|d_1^2 + d_2^2  \right|^2 (2 Y_1 + 4 Y_2 - 4 Y_3 - 2 Y_4)  \right],
\ee
where, again, $X_i \equiv \sum_{\Omega, \bq} V(\Omega, \bq) K_i(\Omega, \bq)$, and $K_i$, $Y_i$ are defined in Eqs.~(\ref{SMeq:K1-K5}) and (\ref{SMeq:Y1-Y4}).

Similarly to the case of CDW fluctuations, the contribution $\delta F_{\Delta}^{(4)}$ always favors the nematic superconducting state, hence, the main result of our paper remains valid for SDW fluctuations as well. The absolute value of this correction, however, is smaller because of a minus sign in front of $Y_3$. The quartic term in $\delta F_{\Delta}^{(2)}$, on the other hand, changes its sign within the specific models for twisted bilayer graphene we considered in Appendices~\ref{SMSec:specificmode1} and \ref{SMSec:specificmode2}. This happens because of an additional minus sign in front of $X_4$. Hence, $\delta F_{\Delta}^{(2)}$ favors chiral state in case of SDW fluctuations. At sufficiently strong fluctuations, however, the fourth-order term in $\delta F_{\Delta}^{(2)}$ is parametrically smaller than $\delta F_{\Delta}^{(4)}$ and thus can be neglected again, justifying our conclusion about the stability of nematic superconductivity.

\end{widetext}

\bibliographystyle{apsrev}

\end{document}